\documentclass[11pt,a4paper,iop]{emulateapj}
\usepackage{epsfig}
\usepackage{amsmath}
\usepackage{natbib}
\usepackage{graphicx}
\usepackage{rotating}
\usepackage{color}

\begin{document}

\title{Being WISE II: Reducing the Influence of Star Formation History on the Mass-to-Light Ratio of Quiescent Galaxies}
\author{Mark A. Norris$^{1,2}$, Glenn Van de Ven$^{1}$, Eva Schinnerer$^{1}$, Robert A. Crain$^{3}$, Sharon Meidt$^{1}$, Brent Groves$^{1,4}$,
Richard G. Bower$^{5}$,  Michelle Furlong$^{5}$, Matthieu Schaller$^{5}$, Joop Schaye$^{6}$, Tom Theuns$^{5}$}
\altaffiltext{1}{Max Planck Institut f\"{u}r Astronomie, K\"{o}nigstuhl 17, D-69117, Heidelberg, Germany}
\altaffiltext{2}{Jeremiah Horrocks Institute, University of Central Lancashire, Preston, Lancashire, PR1 2HE, UK}
\altaffiltext{3}{Astrophysics Research Institute, Liverpool John Moores University, 146 Brownlow Hill, Liverpool, L3 5RF }
\altaffiltext{4}{Research School of Astronomy and Astrophysics, The Australian National University, Cotter Road, Weston Creek, ACT 2611, Australia}
\altaffiltext{5}{Institute for Computational Cosmology, Department of Physics, University of Durham, South Road, Durham, DH1 3LE}
\altaffiltext{6}{Leiden Observatory, Leiden University, PO Box 9513, 2300 RA Leiden, the Netherlands}
\email{mnorris2@uclan.ac.uk}
\date{\today}



\def\Re{{R$_{\rm e}$}}
\def\etal{{\it et al. }}

\begin{abstract}
Stellar population synthesis models can now reproduce the photometry of old stellar systems (age $>$ 2 Gyr) in 
the near-infrared (NIR) bands at 3.4 and 4.6$\mu$m (WISE W1 $\&$ W2 or IRAC 1 $\&$ 2). In this paper we 
derive stellar mass-to-light ratios for these and optical bands, and confirm that the NIR M/L shows 
dramatically reduced sensitivity to both age and metallicity compared to optical bands, and further, that this behavior 
leads to significantly more robust stellar masses for quiescent galaxies with [Fe/H] $\ga$ --0.5
regardless of star formation history (SFH). We then use realistic 
early-type galaxy SFHs and metallicity distributions from the EAGLE simulations of galaxy formation to investigate 
two methods to determine the appropriate M/L for a galaxy:
1) We show that the uncertainties introduced by an unknown SFH can be largely removed using a
spectroscopically inferred luminosity-weighted age and metallicity for the population to select the appropriate single 
stellar population (SSP) equivalent M/L. Using this method, the maximum systematic error due to SFH 
on the M/L of an early-type galaxy is $<$ 4$\%$ at 3.4 $\mu$m and typical uncertainties due to errors in the age and
metallicity create scatter of $\lesssim$13$\%$. The equivalent values for optical bands 
are more than 2-3 times greater, even before considering uncertainties associated with internal dust extinction.
2) We demonstrate that if the EAGLE SFHs and metallicities accurately reproduce the true properties of early-type 
galaxies, the use of an iterative approach to select a mass dependent M/L can provide even more accurate stellar 
masses for early-type galaxies, with typical uncertainties $<$ 9$\%$. 

\end{abstract}

\section{Introduction}

Mass appears to be the principle property of galaxies, and can act as the main parameter determining almost all 
other attributes such as star formation rate \citep[e.g.\ ][]{Peng10,Li11}, assembly history \citep[e.g.\ ][]{Kauffmann03}, 
color \citep[e.g.\ ][]{Peng10}, and morphology \citep[e.g.\ ][]{Bamford09}. When discussing galaxy
mass, it is important to differentiate the different mass components present. Arguably the 
most fundamental property is total dynamical mass which encompasses all baryonic and non-baryonic
(i.e.\ dark matter) components. A close second in importance is the total mass locked up in stars and 
stellar remnants. Finally, there are the generally smaller (at $z=0$) components of gas and dust. 

Determining the exact quantities of each mass component present in a galaxy is a major challenge, with 
the robust determination of stellar mass being notoriously difficult. Fundamentally, all methods to determine 
the stellar mass of galaxies rely on the use of stellar population synthesis (SPS) models to produce estimates 
of the amount of light emitted at some wavelength per unit mass of stars present, i.e.\ the stellar mass-to-light ratio 
(hereafter M/L). At its simplest this can be done assuming a galaxy has a single stellar population (SSP) and 
using a single photometric bandpass. More sophisticated approaches use multiple photometric bands and/or 
spectra in order to attempt to fit more realistic star formation histories and to estimate dust obscuration, 
and thereby to decompose the integrated light into the constituent components emitted by each stellar
population separately (see e.g.\ \citealt{Bell03,Kannappan07,daCunha08,Zibetti09,ATLAS3DXXX,Magris15} for 
various implementations and \citealt{Mitchell13} for a detailed examination of the systematic uncertainties 
present in this approach). In effect this results in the fitting of multiple mass-to-light ratios to each galaxy's
stellar make-up.

However, regardless of the sophistication of the analysis procedure, all approaches to estimate the stellar 
masses of galaxies suffer from several complicating effects including: 1) the strong age dependence of the 
spectral energy distribution of a stellar population that biases the derived properties towards those of the youngest 
population present, 2) the influence of gas and dust absorption and emission, 3) the poorly constrained 
effect of rare but luminous stellar phases such as thermally pulsating asymptotic giant branch stars (TP-AGB),
4) an imperfect knowledge of the star formation history (and the associated metallicity evolution) of the galaxy 
being studied, and 5) the possible variation of the initial mass function (IMF) from galaxy to galaxy or even 
from stellar population to stellar population within galaxies.

The study of the stellar emission of galaxies in the near infra-red has long held promise as a way to reduce
the influence of many of the effects listed above. For example, the NIR M/L ratios of old stellar systems are 
significantly less affected by unconstrained age variations than their optical counterparts \citep[e.g.][]{Meidt14a}.
In addition, in the NIR the metallicity dependence of the mass-to-light ratio is significantly reduced relative
to that seen in the optical \citep[see e.g.][]{WISEI}. The NIR is also significantly less influenced by the presence
of obscuring dust and gas as compared to the optical bands; the V band extinction for example is a factor of 
$\sim$15 times larger than that in the 3.4$\mu$m band assuming a Milky Way-like extinction law \citep{Schlafly11,Yuan13}.
Finally, there is evidence that the influence of poorly constrained stellar phases such as the TP-AGB phase may 
not be as significant as previously thought (for example see \citealt{Gonzalez-Perez14} for a description of 
the differences in predicted galaxy luminosity functions resulting from the use of different TP-AGB prescriptions). 
In particular \citet{McGaugh15} find that all objects in their sample of gas-rich disk galaxies are consistent with having 
the same M/L with very small (0.12 dex) scatter in the IRAC1 \citep{IRAC} band. In this study we
further reduce the confusing influence of TP-AGB stars by focussing on quiescent galaxies, where the influence
from TP-AGB stars is negligible.

Motivated by the dramatic increase in observations of galaxies in the 3.4/3.6$\mu$m bands due to the large 
surveys undertaken by the Spitzer space telescope (e.g.\ SINGS; \citealt{SINGS}, S$^{4}$G; \citealt{S4G}), 
and the all-sky WISE mission \citep[][]{Wright10}, we will quantify the reliability of stellar mass 
estimates derived from this photometry. In particular, we are interested in whether the combination of these
NIR photometric observations with the equally dramatic increase in available spectroscopic observations of
large samples of galaxies (from e.g.\ ATLAS$^{\rm 3D}$; \citealt{ATLAS3DI}, SDSS; \citealt{SDSS}, CALIFA; 
\citealt{CALIFAI}) can be used to significantly improve estimates of stellar mass for older (age $>$ 2 Gyr) 
stellar populations.

This paper is structured as follows. In Section \ref{Sec:SSP_Models} we discuss the stellar population models
used here. In Section \ref{Sec:Effect_of_SFH} we discuss the effect of star formation history (SFH) on mass-to-light
ratios for a range of simple SFHs. In Section \ref{Sec:ml_etgs} we extend this analysis to more realistic SFHs 
for simulated early-type galaxies, and we further examine possible ways to select the appropriate M/L for
early-type galaxies. In Section \ref{Sec:discussion} we discuss the implications of our work, and finally in 
Section \ref{Sec:conclusions} we provide some concluding remarks.

\section{Stellar Population Models}
\label{Sec:SSP_Models}

In this paper we examine the accuracy of the near infrared bands as stellar mass indicators,
in particular when the NIR photometry can be combined with simple spectroscopically derived ages and
metallicities. To do this we make use of the stellar population synthesis models of \citet{Bressan12,Bressan13}.
We use these models because of the current generation of SPS models they were found to most accurately
reproduce the photometry of a large sample of globular clusters and early-type galaxies (age $>$ 2 Gyr)
in Paper I of this series \citep{WISEI}.

We convert the output of the SSP models into predictions for the stellar mass-to-light ratio using the
prescription outlined in Paper I. We use the SSP (i.e.\ single burst) models of \citet{Bressan12,Bressan13} 
with a Chabrier \citep{ChabrierIMF} IMF, which we convert into M/L assuming that the 
fractional mass remaining in stars and remnants over time follows the tracks presented by \citet{Into13} for a 
Kroupa \citep{KroupaIMF} IMF, as the Kroupa IMF mass losses are almost identical to those of a Chabrier IMF 
\citep{Leitner11}.

\begin{figure*} 
   \centering
   \begin{turn}{0}
   \includegraphics[scale=1.1]{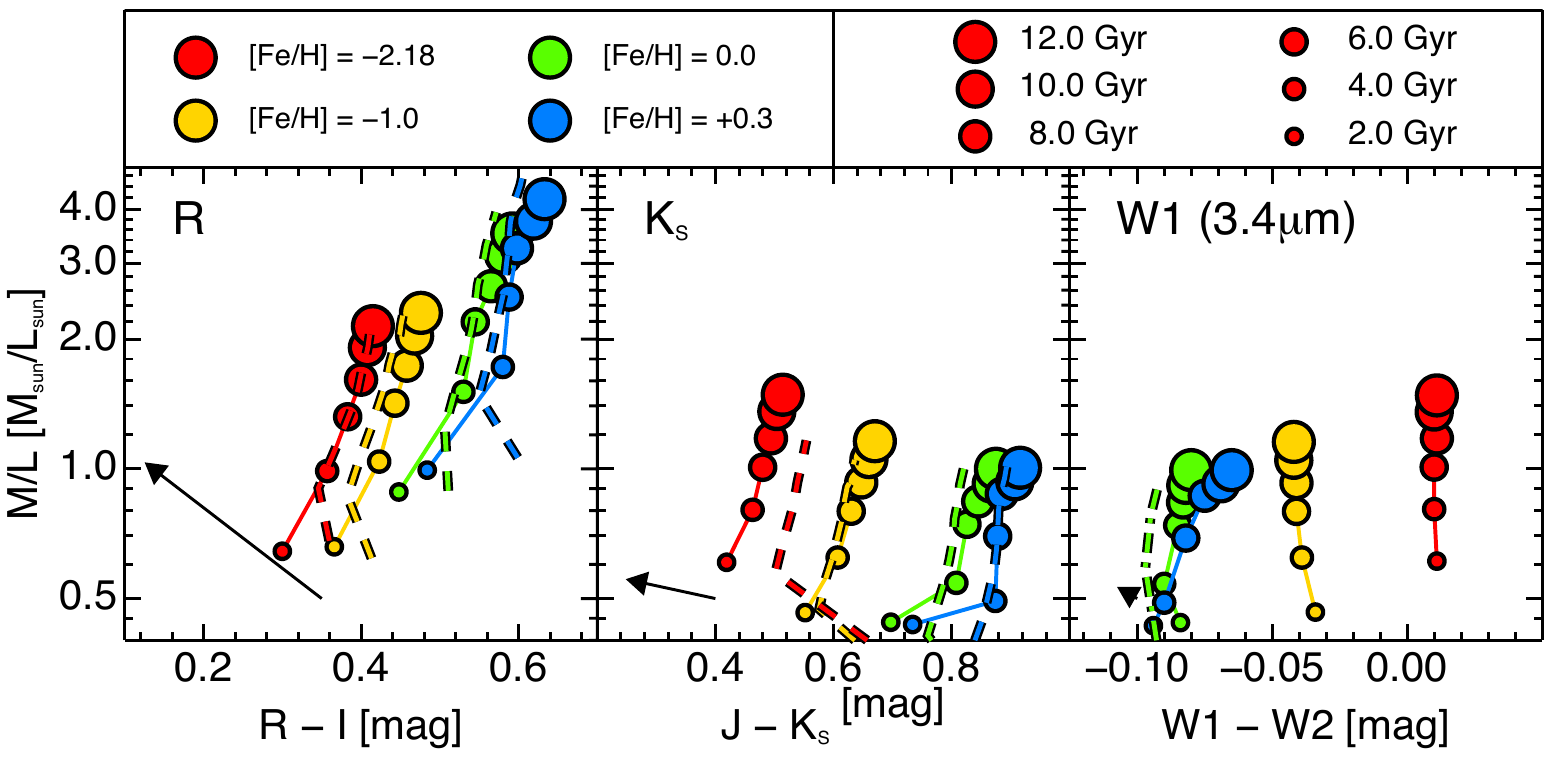}
   \end{turn} 
   \caption{The dependence of the mass-to-light ratio in the R (left panel), K$_{\rm s}$ (central panel), and W1 (right panel) bands 
   vs.\ optical and NIR colour as computed using the \cite{Bressan12,Bressan13} models for a Chabrier IMF (solid
   coloured lines and dots) and additionally using the models of \cite{Maraston05} (R, I, J and K$_{\rm s}$ bands: dashed lines) 
   and \citet{Rock15} (using their Padova based isochrones for the W1 and W2 bands: dot-dashed line) for a Kroupa IMF 
   (as no Chabrier version is available for either model set). 
   The \citet{Maraston05} model predictions are for [Z/H] = [Fe/H] = +0.35, 0.00, --1.35, --2.25, The \citet{Rock15} prediction
   is given for [Fe/H] = 0.0 and ages 3 to 12 Gyr. The different model predictions are in generally good agreement,
   though both the Maraston and R\"ock models extend to slightly lower M/L at younger ages than the Bressan based models
   adopted here.
   The black arrow in each panel demonstrates the effect of an extinction of A$_{\rm V}$ = 1 mag on the implied
   M/L and colour of the SSP assuming the extinction coefficients of \citet{Schlafly11}.
   The reverse colour dependence on metallicity is obvious for the K$_{\rm s}$ and W1 bands. Also of note is the significant
   decrease in sensitivity to age of the M/L for the K and W1 bands as well as the almost total insensitivity to 
   extinction of the W1 band.}
   \label{fig:M_L_Ratio}
\end{figure*}

In Figure \ref{fig:M_L_Ratio} we show the predicted M/L variation with age and metallicity for three photometric 
bands that have been posited as accurate proxies for stellar mass; the NIR K \citep[e.g.\ ][]{KW08} and W1/IRAC1 3.4/3.6$\mu$m 
bands \citep[e.g.][]{Eskew12,Meidt14a,WISEI,Rock15}, and the optical R band 
 \citep[e.g.][]{Kannappan13}. Overplotted 
on each panel (as the dashed lines) where available are the M/L ratios predicted by the models of \cite{Maraston05} 
and the new NIR models of \citet{Rock15} for similar metallicities and a Kroupa IMF for the purpose of comparison. 
The slight variation present between the models may be considered as indicative of the typical theoretical uncertainties 
in the different flavours of SSP models. In particular, as \citet{Rock15} show, the choice of stellar isochrone 
models can affect the derived M/L at old ages by $\sim$10 - 20$\%$. Other potential causes of variation between the models
include differences in the implementation of stellar phases such as TP-AGB stars. 
Nevertheless, the close correspondence between our \citet{Bressan12,Bressan13} derived 
models, and those of \citet{Maraston05} and \citet{Rock15} for the R, K$_{\rm s}$ and W1 bands is heartening, 
indicating broad agreement for those ages where the influence of rare but luminous stellar phases is reduced 
(i.e.\ age $>$ 2 Gyr).

From Figure \ref{fig:M_L_Ratio} it is clear that the sensitivity of the M/L to both age and metallicity is dramatically 
reduced in the W1 band compared to the behavior in the R band. In fact, both the K$_{\rm s}$ and W1 band mass-to-light 
ratios show approximately half the sensitivity to age, and have essentially no dependence on metallicity [Fe/H]  $>$ --1 
dex \citep[see also][]{Rock15}. This insensitivity to metallicity $>$ --1 dex has particularly useful implications for the study 
of massive galaxies, since these generally have metallicities in this range. Unfortunately, in contrast to the behaviour 
in the NIR, the R band M/L approximately doubles when the metallicity is increased from -1 to 0 dex at fixed age and 
only displays insensitivity to metallicity at lower metallicities than generally observed in galaxies.

Also plotted in each panel of Figure \ref{fig:M_L_Ratio} is an arrow denoting the effect on the measured colors and 
M/L ratio if an extinction of 1 magnitude A$_{\rm V}$ is present but uncorrected. It is clear that the effect of extinction,
which is already significantly reduced in the K$_{\rm s}$ band, is further reduced by a factor of two when moving to the W1 band.

\begin{figure*}
\centering
\begin{tabular}{c}
   \includegraphics[scale=1.0]{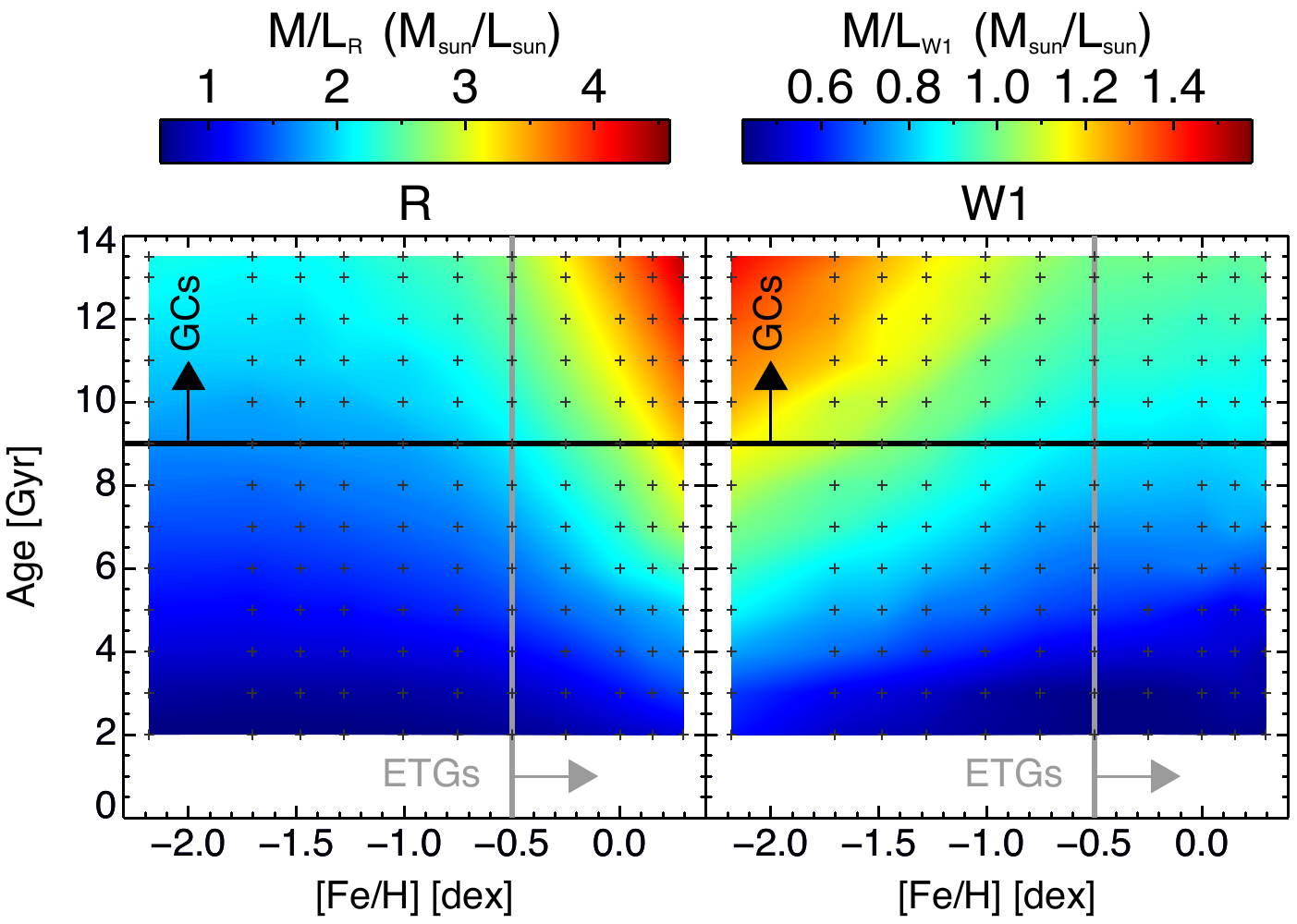} \\
   \includegraphics[scale=1.0]{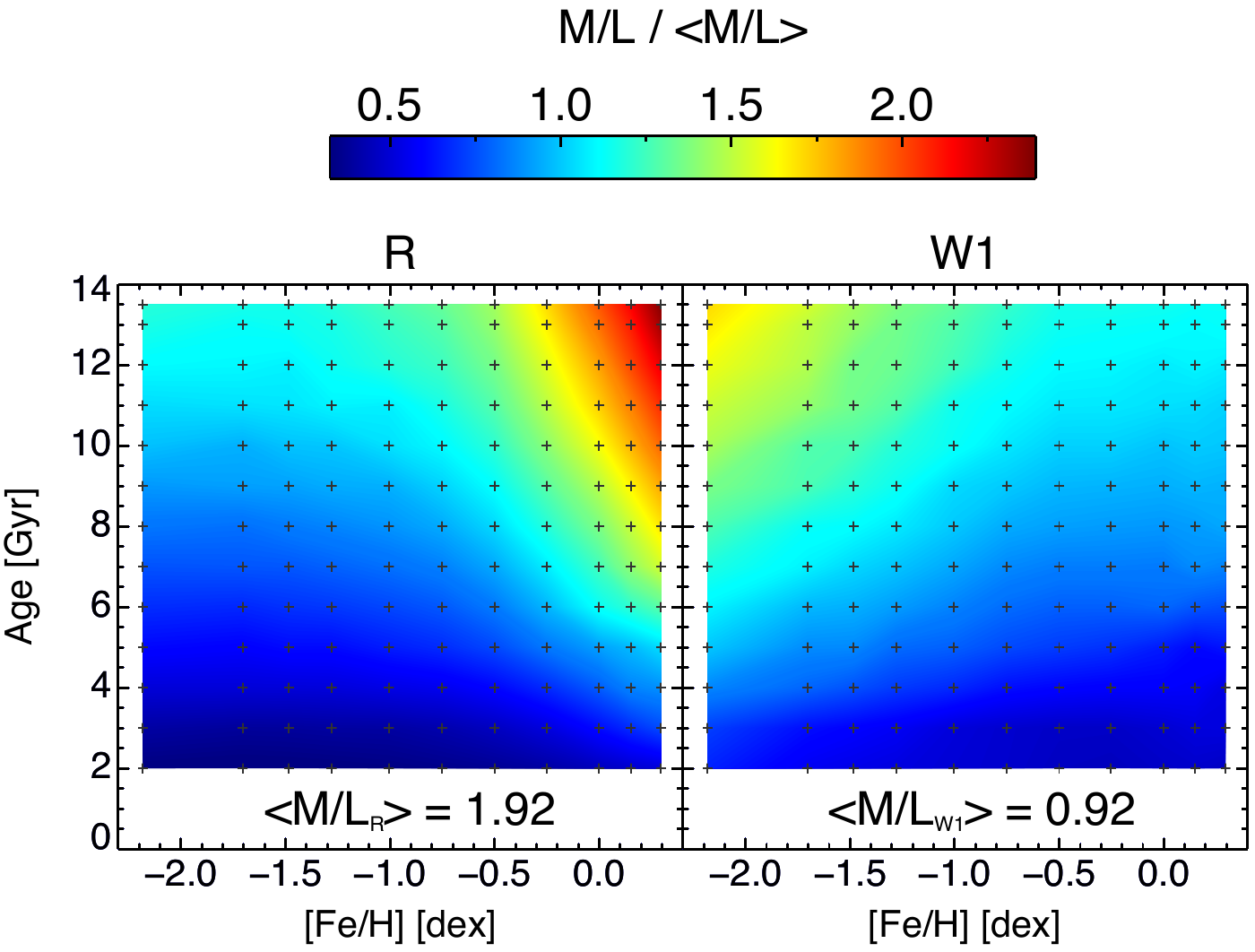} 
\end{tabular}
\caption{\textbf{Upper Panels:} The dependence of the mass-to-light ratio in the R (left) and W1 (right) bands on the age and metallicity 
   of SSPs, as derived from the models of \cite{Bressan12,Bressan13} for a Chabrier IMF. The crosses indicate
   the exact SSP metallicity and age combinations available, which were interpolated to produce the maps.
   Also overplotted are two lines which indicate the ages and metallicities displayed by globular clusters (typical ages $>$ 9 Gyr,
   all metallicities) and early-type galaxies (metallicities $>$ --0.5 dex, ages $>$ 2 Gyr). The very weak 
   dependence of M/L on metallicity for [Fe/H] $>$ --0.5 in the W1 band is apparent at all ages. The opposite
   is true for the R band, where the dependency on [Fe/H] is weakest at lower metallicities.\\
   \textbf{Lower Panels:} The relative change in M/L for the R (left) and W1 (right) bands. Each panel has been rescaled by the
   median of the mass-to-light ratios in each band (1.92 in the R, and 0.92 in W1). The
   relative change in M/L for the W1 band is a maximum of a factor of 2 for ages and metallicities typical of galaxies,
   while for the R band it is around a factor of 5.}
\label{fig:M_L_Grid}
\end{figure*}

The upper panels of Figure \ref{fig:M_L_Grid} display a smoothed interpolated grid of SSP models, 
showing the dependence of M/L in the R and W1 bands on age and metallicity. The almost total independence 
of the mass-to-light ratio of W1 on metallicity for metallicities $>$ --1 dex is now even more obvious. As 
most large galaxies have metallicities in excess of this value, this observation has the potential to be 
extremely useful when deriving stellar masses for such systems. Additionally, as the lower panels of 
Figure \ref{fig:M_L_Grid} illustrate more clearly (by normalising each M/L grid to the median of the M/L 
in each band), for metallicities typical of galaxies the change in M/L when increasing age from 2 to 13.5 Gyr 
is about a factor of 4 in the R band, but less than a factor of 2 in the W1 band (and a very similar factor 
in K$_{\rm s}$). This relative insensitivity to age implies that the M/L of NIR bands should also benefit from reduced 
sensitivity to star formation history.

\section{The Effect of Star Formation History on Mass-to-Light Ratios}
\label{Sec:Effect_of_SFH}

In the preceding section we demonstrated that M/L ratios of the W1 band have several significant
advantages over optical bands and can be useful stellar mass indicators for the simplest stellar populations. 
However, we did this using single stellar population models. In reality most galaxies consist of composite 
populations formed in many individual star forming events. As such the stellar populations 
of galaxies are a mixture of stars with a potentially wide range of ages and metallicities. In the 
real world the usefulness of single mass-to-light ratios derived from SSPs is therefore not obvious.

There are several potential ways to address the problem of the composite nature of the stellar populations 
of galaxies affecting the measurement of stellar mass. One widely used approach is to attempt to 
decompose the stellar population of a galaxy into its individual components, or more accurately into a series 
of pseudo-SSP populations that can then be individually multiplied by their appropriate M/L ratio to give the
total stellar mass \citep[see e.g.][]{ATLAS3DXXX,Norris15}. This approach requires significantly higher quality 
data that is not available for most current studies and is fraught with degeneracies in the star formation 
history parameters that can potentially lead to non-unique solutions.

Another approach is to attempt to marginalise over the uncertainty in the SFH by comparing the observed
SED/spectra with models that attempt to cover a wide range of SFHs \citep{Kauffmann03a,Zibetti09}. This
method suffers from some of the same problems associated with approach 1, as well as uncertainties regarding
how well the library of SFHs reproduce those of real galaxies.

A final potential approach is hinted at by the relative insensitivity of the NIR bands to both age and 
metallicity, as seen in Figures \ref{fig:M_L_Ratio} $\&$ \ref{fig:M_L_Grid}. In this approach we first examine 
the actual sensitivity of the M/L in various bands to changes in star formation history and attempt to quantify 
its importance.

To this end we have computed a range of composite stellar populations (CSPs) using the SSP models and our
derived M/L ratios described previously. These models come in two forms, simplistic SFHs designed to illuminate
general behavior, and a second set of ``realistic" SFHs derived from the EAGLE galaxy formation simulations 
\citep{EAGLEI,EAGLEII}. We use simulated galaxies, because no large sample of well constrained (i.e.\ with 
good age and metallicity resolution) SFHs exist for observed galaxies, particularly for early-type galaxies. Principally 
this is because very few massive early-type galaxies exist within distances (i.e.\ within $\sim$10 Mpc) for which it 
is possible to resolve their stellar populations into individual stars. This forces the analysis of early-type galaxies 
to be done using integrated properties, which inevitably leads to significantly larger uncertainties.

In the case of the simple SFHs, we create our CSPs by constructing a SFH from a series of pseudo-bursts 
of duration 100 Myr, i.e.\ all stars formed in each 100 Myr bin are all assigned the same age and metallicity.  

For the realistic star formation histories from the EAGLE simulations we wish to adequately reproduce the age 
and metallicity combinations present in galaxy populations, without having to analyse each star particle individually 
(the largest galaxies have $>$ 500,000 particles). We therefore bin the star particles by logarithmic age and 
metallicity to provide 41 values of age, with bin centres from 133 Myr to 13.3 Gyr, and 20 of metallicity from --4 to 
+1 dex. We then convert these composite SFHs into the combined M/L for the whole population at $z=0$, and 
determine the R band luminosity-weighted age and metallicity of the composite population. 
In the later analysis we assume axiomatically these values are the same as the luminosity-weighted age and 
metallicity that would be derived using spectroscopic methods, such as line index measurements or full spectral 
fitting as used in, for example, the ATLAS$^{\rm3D}$ \citep{Kuntschner10,ATLAS3DXXX} survey. With the 
luminosity-weighted age and metallicity we then determine the SSP-equivalent M/L for the population and compare 
this to the true integrated M/L of the simulated galaxy. In order to achieve a more representative comparison with the 
observations, where the uncertainties on luminosity-weighted age and metallicity can be considerable, we run 100 
Monte Carlo re-simulations of the data where we add appropriate measurement errors to the luminosity-weighted 
age and metallicity and see what effect this has on the mismatch between the true and SSP M/L. 

\begin{figure} 
   \centering
   \begin{turn}{0}
   \includegraphics[scale=0.9]{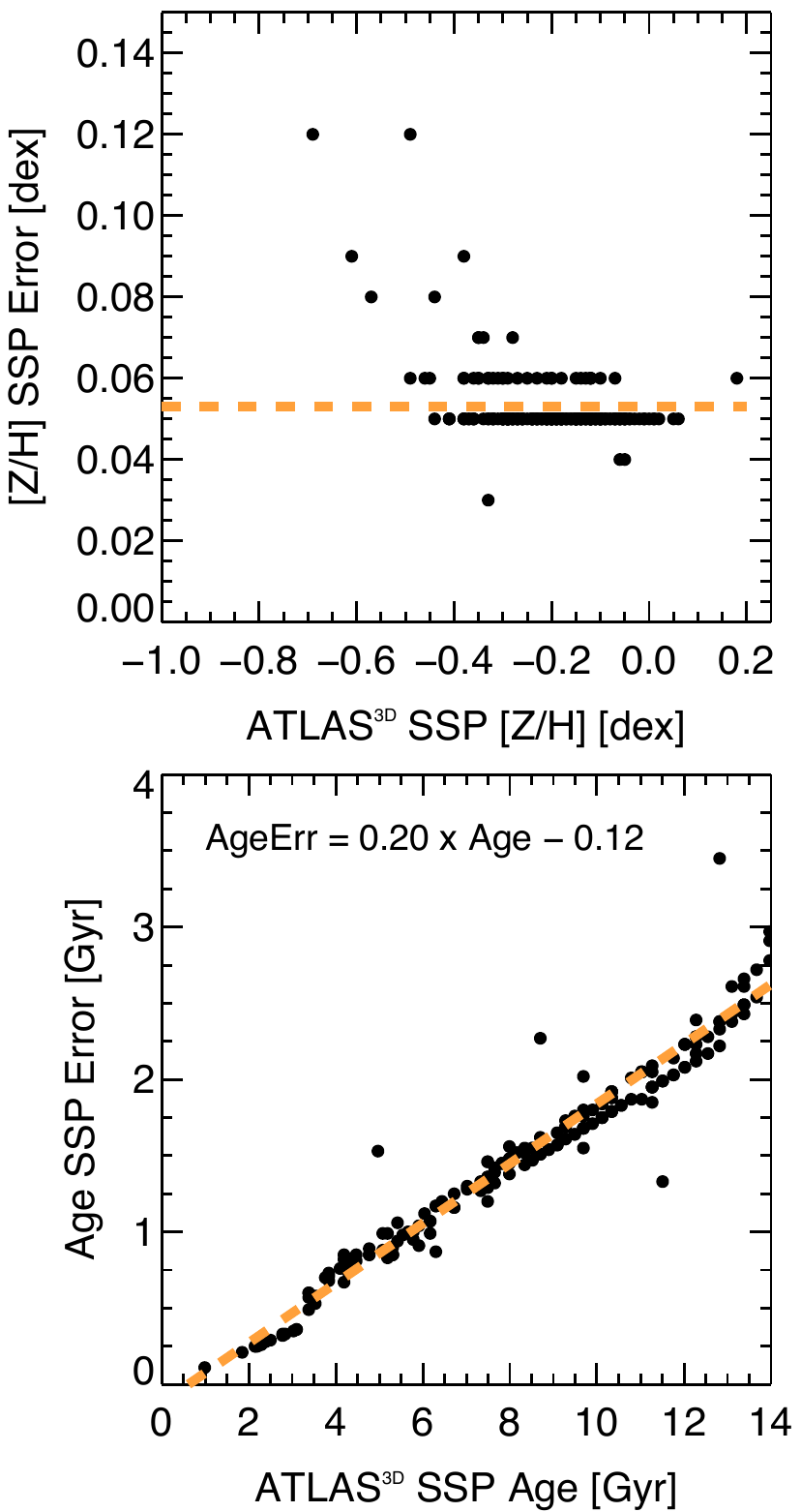}
   \end{turn} 
   \caption{The correlation of inferred SSP age and metallicity with their associated errors for the ATLAS$^{\rm 3D}$
   galaxy sample as measured by \cite{ATLAS3DXXX}. Here we examine only those galaxies whose quality flag 
   indicated that the data quality was good (flag = 1) and whose age was found to be less than 14 Gyr. The dashed 
   line in the lower panel is a linear fit to the data, the corresponding best-fit parameters are shown in the top left of 
   the panel.  As there is no significant trend in the metallicity vs.\ metallicity error plot, no fit is made and only the 
   mean value of the metallicity error is indicated in the upper panel by the dashed line.  }
   \label{fig:ATLAS3D_Errs}
\end{figure}

The uncertainties on the age and metallicity used in the Monte Carlo analysis are chosen to match those reported 
by the ATLAS$^{\rm 3D}$ survey \citep{ATLAS3DXXX}. We do this because we make the working assumption
that the luminosity-weighted stellar population ages and metallicities of our simulated SFHs are equivalent to the SSP-equivalent 
ages that would be measured using the spectroscopic techniques employed by for example the ATLAS$^{\rm 3D}$ survey. 
Note however, that while this is a very good assumption for metallicity, it is not always the case for age 
\citep{Trager&Sommerville2009}, SSP ages greater than around 4 Gyr do tend to track lumionsity-weighted age quite well,
but at younger ages the correspondence becomes much weaker. Our choice to use realistic errors on the derived
age and metallicity is important, because, as expected and 
Figure \ref{fig:ATLAS3D_Errs} demonstrates, the measurement uncertainty on the SSP equivalent age is a function 
of age (but approximately a constant fraction, 15-20$\%$, of the age). This is due to the fact that the primary age 
sensitive absorption feature used by the ATLAS$^{\rm 3D}$ survey to determine the age of a stellar population, 
H$\beta$, evolves most rapidly at younger ages. Therefore, for a fixed signal-to-noise ratio (and hence uncertainty 
on the line strength index), the absolute uncertainty on the age is lower for younger ages. In practice when carrying 
out the Monte Carlo analysis, the luminosity-weighted age (i.e.\ SSP equivalent) of the full SFH is calculated, then the 
error on that age is determined following the relation shown in Figure \ref{fig:ATLAS3D_Errs}. This means that the 
uncertainty on an SSP equivalent age drops from around $\pm$2 Gyr at 12 Gyr, to around $\pm$0.25 Gyr at 2 Gyr. In 
contrast to the situation for the measured SSP equivalent ages, the SSP equivalent metallicities from the ATLAS$^{\rm{3D}}$ 
survey are found to have errors that are approximately independent of metallicity (and age). Therefore a fixed 
uncertainty of 0.052 dex is applied to the luminosity-weighted metallicity during the Monte Carlo analysis.

\subsection{Single Burst Star Formation Histories}

\begin{figure} 
   \centering
   \begin{turn}{0}
   \includegraphics[scale=0.9]{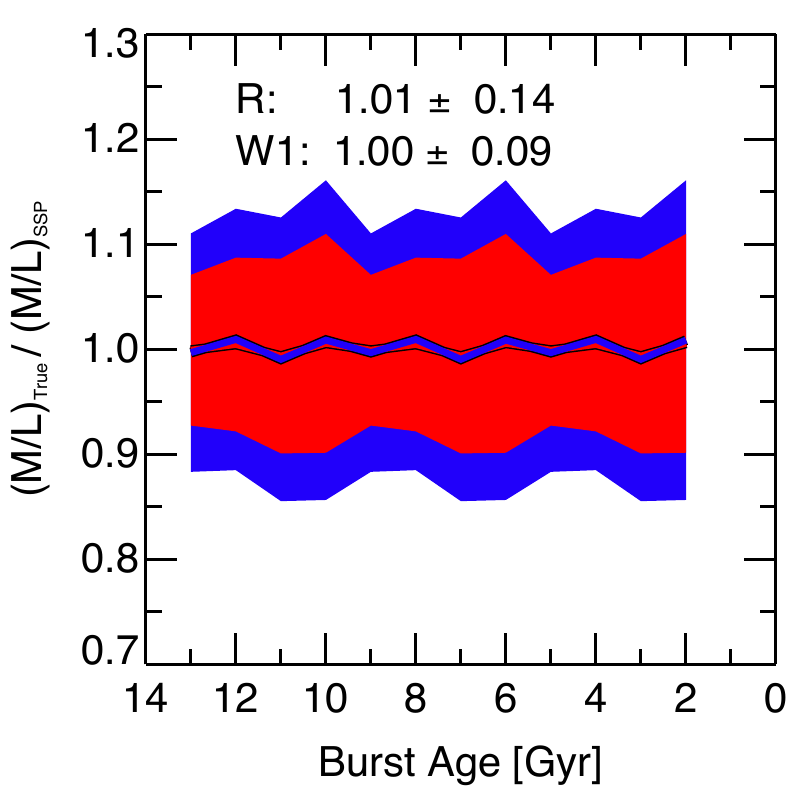}
   \end{turn} 
   \caption{The ratio of the True M/L to the luminosity-weighted SSP-equivalent M/L for single burst 
   models with age between 13 and 2 Gyr. The red and blue lines (with black edges for 
   visibility) show the median of the ratios for the W1 and R band respectively. As expected for such 
   simple models, the SSP M/L reproduces the true M/L for all ages (i.e. the ratio is very close to unity
   for all ages and both bands). The shaded regions indicate the 1-$\sigma$ scatter ($\sim$14 and 9$\%$ 
   for R and W1 respectively) based on 500 Monte Carlo simulations for each burst where the scatter 
   is produced by assuming realistic observational uncertainties on the luminosity-weighted age and 
   metallicity used to determine the SSP M/L (see Section \ref{Sec:Effect_of_SFH}). Errors on the
   measured luminosity are assumed to be negligible for this analysis.   
   }
   \label{fig:GC_ML}
\end{figure}

The simplest possible SFH is a single burst population with a single metallicity, such as those often assumed for compact 
stellar systems such as globular clusters (see e.g.\ \citealt{Kuntschner02,Puzia05,Norris08}). Figure \ref{fig:GC_ML} 
shows the behaviour of the true M/L to SSP M/L ratio as a function of the age of the single burst, for both the 
R and W1 bands. As expected for such a simple SFH, both the R and W1 SSP M/L ratios accurately reproduce 
the true M/L in the sense that the ratio M/L$_{\rm True}$ to M/L$_{\rm SSP}$ is indistinguishable from unity. Furthermore, 
this figure also demonstrates that the decreased sensitivity of W1 to age described in Section \ref{Sec:SSP_Models} 
leads to a corresponding decrease in the scatter of the M/L$_{\rm True}$ to M/L$_{\rm SSP}$  ratio, which is 
reduced in the W1 band by around 50$\%$ compared to R, to a statistical uncertainty of less than 10$\%$ 
at all ages. Obviously, it is not surprising that we are able to recover the M/L ratio of an input SSP accurately. 
However, this plot does demonstrate the absolute minimum uncertainty obtainable on the derived M/L 
ratio under the assumption of a true SSP, in the absence of dust, and with only random uncertainties
on the age and metallicity described in the previous section.

\subsection{Constant Star Formation Histories}

\begin{figure*} 
   \centering
   \begin{turn}{0}
   \includegraphics[scale=0.9]{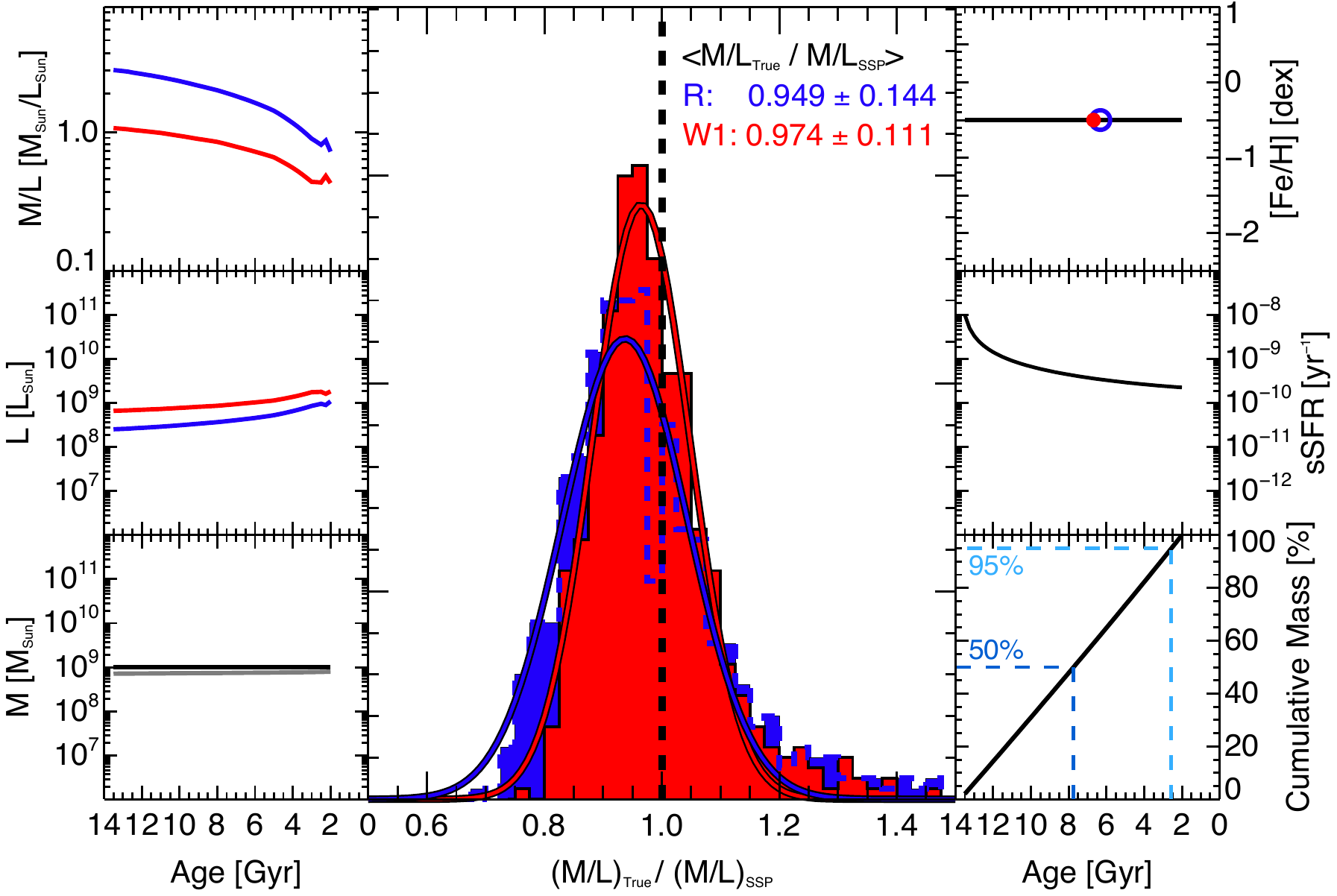}
   \end{turn} 
   \caption{A toy model of a constant SFH from 14 to 2 Gyr at constant metallicity.
   $\bf{Left\,Lower\,Panel:}$ Mass of stars formed per 100 Myr (black line), mass of stars and remnants 
   remaining at $z=0$ (grey line). $\bf{Left\,Middle\,Panel:}$ The luminosity from each 100 Myr bin at $z=0$
   for the R (blue) and W1 (red) bands. $\bf{Left\,Upper\,Panel:}$ The M/L ratio of each 100 Myr bin at $z=0$
   for both bands. $\bf{Right\,Lower\,Panel:}$ The cumulative mass fraction against time for the SFH. The 
   dashed lines show the age when 50 and 95$\%$ of the stellar mass was in place. $\bf{Right\,Middle\,Panel:}$ 
   The specific star formation rate (stellar mass formed per bin divided by total stellar mass) of the SFH. 
   $\bf{Right\,Upper\,Panel:}$ The metallicity distribution of the SFH (black line). The blue and red circles are, 
   respectively, the R and W1 band luminosity-weighted age and metallicity of the full population.   
   $\bf{Central\,Panel:}$ Histogram of the true M/L divided by the M/L of the SSP with age 
   and metallicity equal to the R-band luminosity-weighted average age and metallicity of the population.
   The blue histogram is for the R band and the red for the W1 respectively. In both cases the the
   age and metallicity used to derive the SSP M/L is the R band luminosity-weighted age and metallicity, as
   this is closest to what is used observationally, and the difference between the two is negligible anyway (see
   the red and blue circles in the top right panel). The spread arises because we have carried out 100 
   Monte Carlo simulations where the true luminosity-weighted age and metallicity has been altered by the 
   typical measurement uncertainties described in Section \ref{Sec:Effect_of_SFH}. The red and blue curves 
   are the best fit Gaussians to the distributions and highlight the slightly asymmetric nature of the distributions
   caused by the fact that under- and over-estimated ages do not lead to equal changes in M/L due to the
   evolution of M/L with time.
   The text in the upper right shows the median ratio of the True M/L to the SSP M/L along with the
   1-$\sigma$ scatter in each distribution.}
   \label{fig:ContinuousSFH}
\end{figure*}

The next step towards a more realistic SFH for a galaxy is to include populations with a range of ages. 
Figure \ref{fig:ContinuousSFH} shows a constant SFH, with stars being formed at a constant rate from 14 to 2 
Gyr with constant metallicity. In this figure we can see that for the constant SFH the true M/L to SSP M/L 
ratio is not unity for either filter, but that the W1 band does do a better job of matching the true M/L and still with 
reduced scatter relative to the R band. Repeating the procedure for different metallicities yields similar results: 
in all cases the W1 band SSP M/L is significantly closer to the true value with lower scatter than for the R band. 
A further feature common to all constant SFHs is that it is clear that the spread in the derived M/L$_{\rm True}$ to 
M/L$_{\rm SSP}$ ratio is slightly asymmetric, with the median ratio indicating that in most cases the Monte Carlo 
simulations show that the SSP M/L overestimates the true M/L, while the tail indicates that a significant fraction of 
SSP models underestimate the true M/L. Part of the asymmetry reflects the asymmetry in the luminosity evolution 
of a stellar population; an error towards younger ages has a larger effect on the M/L than an equal error (in linear 
age) towards older ages.

Another step towards increasing realism is to include metallicity evolution. Due to enrichment of the inter stellar 
medium by supernovae and stellar mass loss, subsequent generations of stars are typically more metal-rich than 
than those that preceded them, although the accretion of fresh, low-metallicity gas from the inter galactic medium 
can complicate this picture. We have experimented with a range of constant and exponentially declining SFHs
(with $\tau$ of 1, 2 and 4 Gyr ), with variable metal enrichment (constant or monotonically increasing). 
We find that in common with the cases above, the W1 band SSP M/L is almost always 
closer to the true M/L and always has lower scatter than the R band SSP M/L. The few cases where
the R band does provide a better match to the true M/L result from SFHs which, while not technically unphysical,
are not seen in the real Universe, i.e. an exponentially declining  star formation history with a monotonically smoothly 
increasing metallicity that takes 14 Gyr to get from --2 dex to solar. 
However, a common occurrence with the 
models is for a slight, possibly systematic, mismatch between the SSP and the true M/L. In order to examine this 
point further it is necessary to examine more realistic SFHs.

\subsection{``Realistic" Star Formation Histories}
\label{RSFH}

In this section we make use of the EAGLE simulations to examine the effect of more realistic SFHs on the 
derived M/L. The EAGLE simulations comprise a set of hydrodynamical simulations of cosmologically
representative volumes of the Universe with prescriptions for various forms of stellar and AGN feedback. 
Here we use the largest (in terms of volume and particle number) EAGLE simulation, 
i.e. model Ref-L100N1504, which consists of a 100 Mpc$^3$ volume, which yields more than 325,000 galaxies 
with total stellar mass greater than 10$^{9.5}$ M$_\odot$ at $z=0$. This large sample of simulated galaxies 
broadly reproduces many observed properties of the galaxy population, including; the stellar mass function, 
passive galaxy fraction, and most critically for us, the mass-metallicity relation \citep{EAGLEI}. In addition, 
EAGLE broadly reproduces the observed evolution of the galaxy mass function \citep{Furlong15}
and the observed bi-modality of galaxy colours \citep{Trayford15}. Furthermore, the simulations have the 
benefit that they track the formation of every star particle (each of which has a mass of approximately 10$^6$ M$_\odot$ 
and can be treated as an SSP), whether it formed initially within the most massive progenitor of the $z$=0 galaxy, 
or within an initially separate galaxy that was accreted by $z$=0.  This means that the derived SFHs are 
much closer to those of real early-type galaxies (which are believed to form through a two phase build 
up; see e.g.\ \citealt[][]{Oser10}), than the simple toy models we have  discussed so far.

\begin{figure*} 
   \centering
   \begin{turn}{0}
   \includegraphics[scale=0.9]{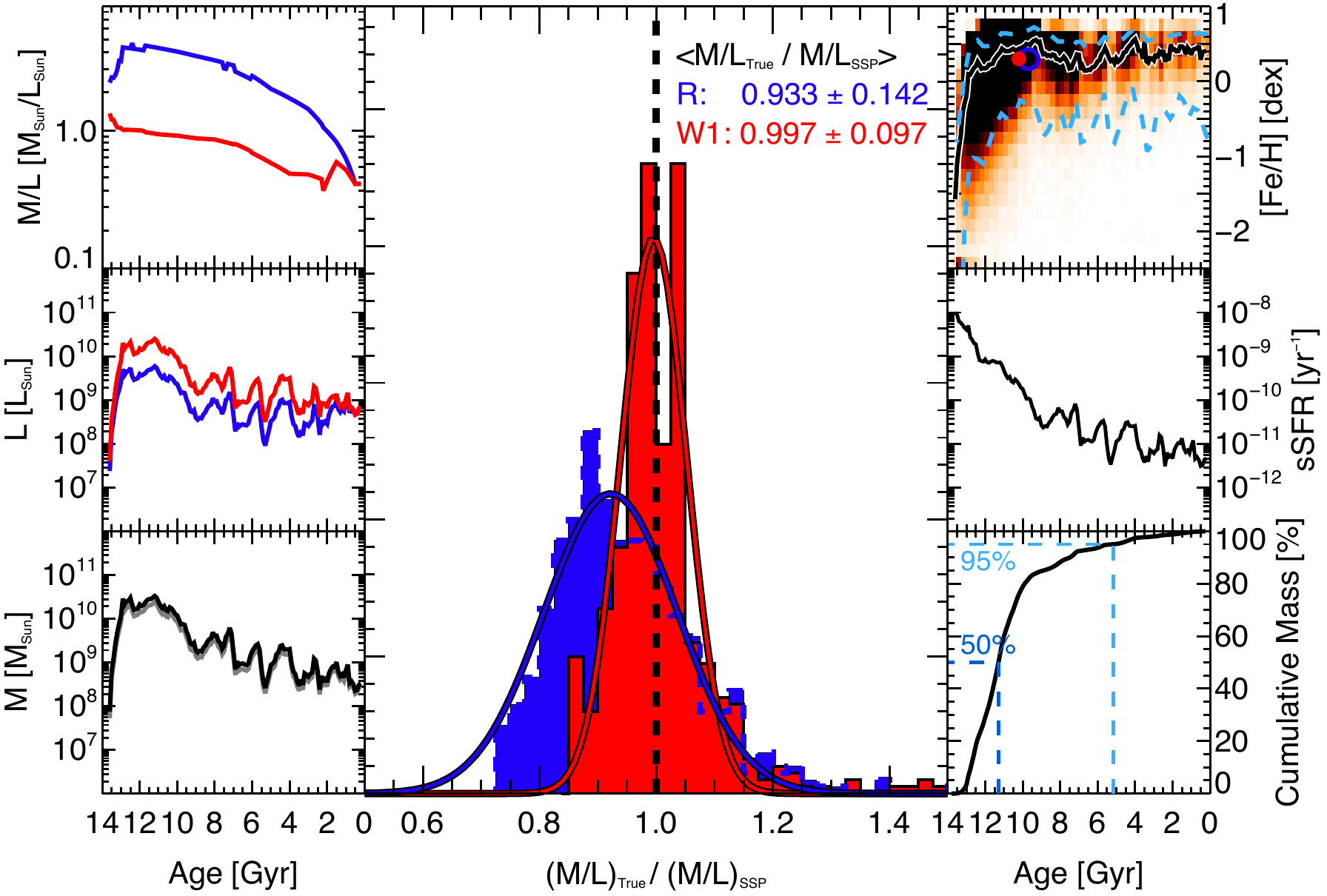}
   \end{turn} 
   \caption{Star formation history and resulting M/L$_{\rm True}$ to M/L$_{\rm SSP}$ ratio for a massive
   (M$_\star$ = 6$\times$10$^{\rm 11}$M$_\odot$) early-type galaxy from the EAGLE simulation. The format 
   is the same as in Figure \ref{fig:ContinuousSFH} except that in the age vs metallicity panel the background
   colour scale shows the age and metallicity distribution of all stellar particles which end up in the galaxy in
   $z$=0. The solid black and white line in this case shows the mean metallicity in 250 Myr wide bins of age, 
   while the cyan dashed lines show the 10 and 90$\%$ percentiles of the metallicity distribution. From
   this distribution it is clear that the metallicity evolution of the galaxy is complex, it is not simply a monotonically rising
   function of the age, it is also clearly heavily skewed towards higher values of metallicity, at all ages. At various points 
   accreted gas or stars from smaller galaxies lead to decreasing average metallicity for stars of that particular age bin 
   relative to those around it. At other times in-situ star formation rapidly drives the metallicity back up to the more typical 
   value of $\sim$ 0.5 dex.
    }
   \label{fig:MassiveGalaxy}
\end{figure*}

Figure \ref{fig:MassiveGalaxy} shows the SFH and metallicity enrichment history of one such simulated galaxy, 
in this case one of the more massive galaxies in the simulation with M$_\star$ = 6 $\times$ 10$^{\rm 11}$ M$_\odot$. 
From examination of the bursty nature of the SFH (bottom left panel) and complex metallicity 
evolution (upper right panel), it is immediately obvious that the evolution of this galaxy is much more realistic 
than the toy models examined previously. The metallicity evolution in particular captures both periods of rapid 
increase, during the initial burst of star formation, and then periods of slight decline due to the accretion of 
lower-metallicity dwarf galaxies or pristine gas from the inter galactic medium. This galaxy also suggests that 
even the most massive early-types could still be actively forming some stars at $z=0$. In fact, the rate of star 
formation and even the sSFR at $z=0$ can be similar to that of the Milky Way. However, 
compared to the vast bulk of stars already formed, the recently formed stellar mass is negligible. In this 
simulated galaxy only 5$\%$ of the stars were formed in the last 5.5 Gyr. It is also notable that despite the
more complex star formation and metallicity enrichment history (compared to the toy model seen in
Figure \ref{fig:ContinuousSFH}), the offset and scatter between true and SSP M/L is still significantly smaller 
in the W1 band for this galaxy.

As described in Section \ref{Sec:Effect_of_SFH} we bin the EAGLE SFHs into bins of age and metallicity
in order to speed up the analysis. We have examined a subset of galaxies using both the binning prescription 
and one where we analyse each star particle individually, this analysis confirmed that our binning procedure 
leads to no significant differences when compared to analysing the full star particle sample. However, this is 
not the case if the data is binned only by age i.e. if the average metallicity is calculated at each timestep (e.g. 
the black solid line in the age-metallicity panel of Figure \ref{fig:MassiveGalaxy}), in such a binning scheme 
the influence of lower metallicity stars at younger ages (from accreted dwarfs) is suppressed.

\section{Mass-to-Light Ratios of Simulated Quiescent Galaxies}
\label{Sec:ml_etgs}
\subsection{Selecting Quiescent Galaxies}
\label{Sec:etg_selection}

We are interested primarily in SFHs that lead to early-type/quiescent galaxies, because the light output at 3.4/3.6$\mu$m 
for such galaxies is unaffected by significant dust emission that would bias the M/L determination, and the 
maximum contribution due to non-stellar emission is 15$\%$ for such galaxies \citep{Querejeta15}. 
We first limit our selection to galaxies with total stellar mass $>$ 10$^{9.5}$ M$_\odot$ at $z=0$ 
to ensure adequate sampling of the ISM and hence to reduce resolution-dependent stochasticity in the star 
formation rate \citep{Furlong15}. We choose to use the specific star formation rate (sSFR) vs.\ stellar mass 
space as an quiescent galaxy selector as opposed to a more typical color-magnitude or color-stellar mass 
selection, because this measure is independent of the choice of post-processing done to the EAGLE simulations 
to convert their stellar masses into luminosity. In other words we avoid any problems 
caused by there being a mismatch between the models used by the EAGLE simulations to convert mass-to-light, 
and our approach. In practice these selections also ensure that none of the galaxies selected form more than 
5$\%$ of their stars within the last 2 Gyr. Such a limit also has the positive benefit for real observed galaxies of 
limiting the uncertainty caused by highly luminous AGB phases which are most problematic for young ages and 
are not well constrained in the SSP models used to determine the M/L.

\begin{figure*} 
   \centering
   \begin{turn}{0}
   \includegraphics[scale=1.10]{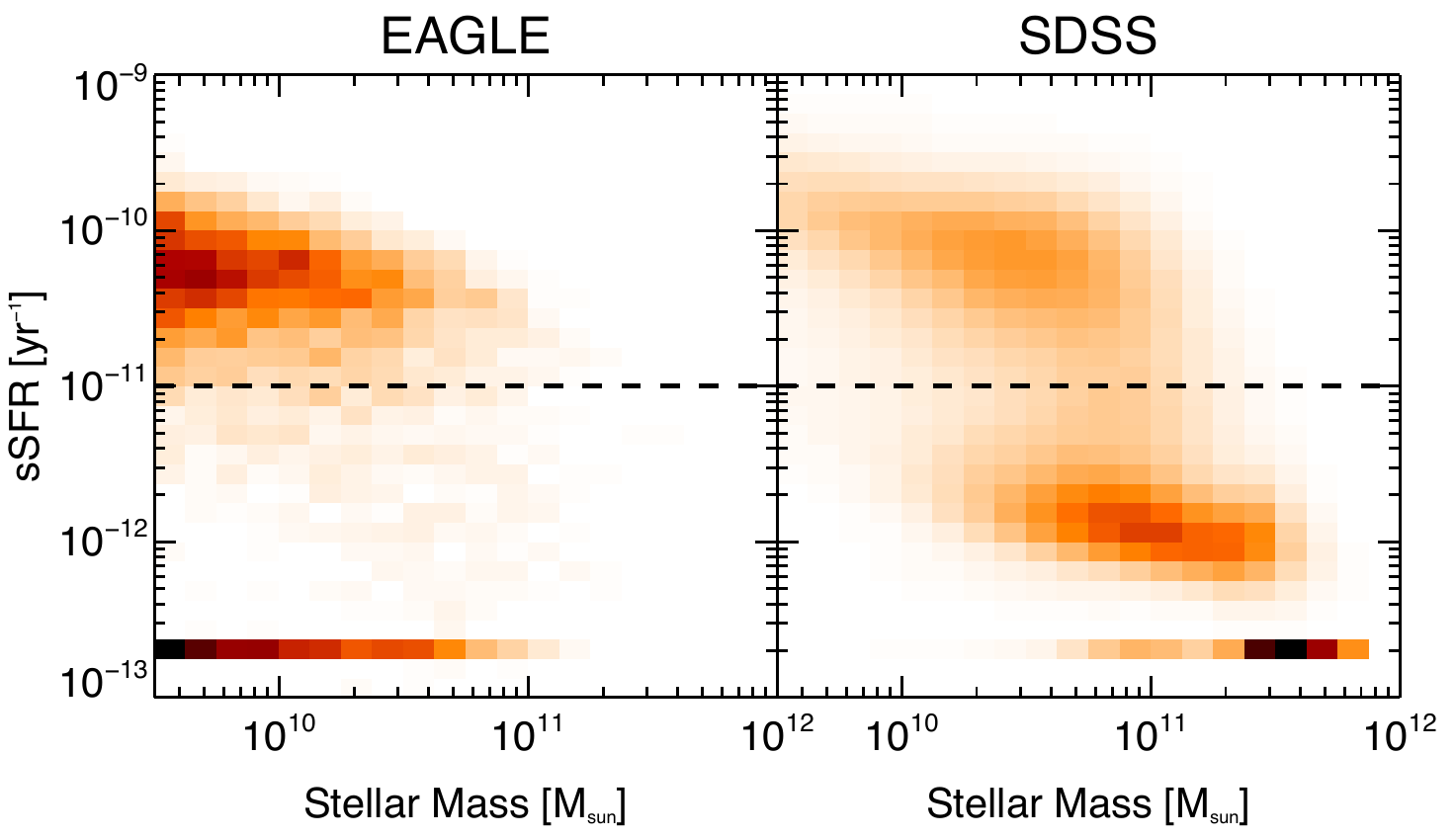}
   \end{turn} 
   \caption{\textbf{Left Panel:} The specific star formation rate vs.\ stellar mass for simulated
   galaxies from the EAGLE projects 100 Mpc$^3$ volume \citep{EAGLEI,EAGLEII} for all 
   galaxies with M$_\star$ $>$ 10$^{9.5}$M$_\odot$. All simulated galaxies which have sSFR $<$ 
   10$^{-13}$ yr$^{-1}$ (which includes many with sSFR=0, i.e.\ no star formation at all in 
   the last time step) are plotted with 2$\times$10$^{-13}$ yr$^{-1}$ for clarity.
   \textbf{Right Panel:} The specific star formation rate vs.\ stellar mass for the SDSS sample 
   from the MPA/JHU VAGC \citep{Kauffmann03a,Brinchmann04,Salim07}. As in the left panel
   all galaxies which have sSFR $<$ 10$^{-13}$ yr$^{-1}$ are plotted with 2$\times$10$^{-13}$ 
   yr$^{-1}$ for clarity. The red sequence, green valley, and blue cloud are all clearly visible in this
   parameter space, with the red sequence particularly prominent as the over density of objects
   with sSFR $\sim$ 10$^{-12}$ yr$^{-1}$. The apparent difference between the EAGLE and
   SDSS distributions is due to the difference between the \textit{volume} limited EAGLE 
   pseudo-survey and the \textit{magnitude} limited SDSS survey.
   The dashed line in each panel shows our chosen sSFR limit used to select quiescent/early-type
   galaxies.
   }
   \label{fig:SDSS_sSFR}
\end{figure*}

The left panel of Figure \ref{fig:SDSS_sSFR} shows the sSFR for the full sample of EAGLE galaxies from the 
100 Mpc$^3$ volume, while the right panel shows the observed distribution from the SDSS 
\citep{Kauffmann03a,Brinchmann04,Salim07}. The comparison between observed and simulated galaxies is 
not entirely fair, as the simulated galaxies are able to measure sSFRs that are 
significantly lower (the minimum sSFR is effectively set by the minimum particle mass) than the observed 
galaxies, where observational effects mean that either upper limits are all that can be measured, or the sSFR 
is fixed to be zero. Furthermore, there is a fundamental difference between the EAGLE sample and 
that of SDSS, the former is $\textit{volume}$ limited, while the latter is a $\textit{magnitude}$ limited. Regardless, 
it is clear that simulated galaxies exist with approximately the correct combinations of sSFR and stellar mass 
to be safely selected as early-type/quiescent galaxies. Given that the volume-limited sample (if large enough to reduce 
cosmic variance) is the truer representation of the real Universe we therefore select all EAGLE simulated galaxies 
which have sSFR $<$ 10$^{-11}$ yr$^{-1}$ as our simulated quiescent galaxy sample. This selection yields a
total sample of 2185 simulated galaxies.

\subsection{Estimating M/L Using SSP-Equivalent Ages and Metallicities}

\begin{figure*} 
   \centering
   \begin{turn}{0}
   \includegraphics[scale=1.0]{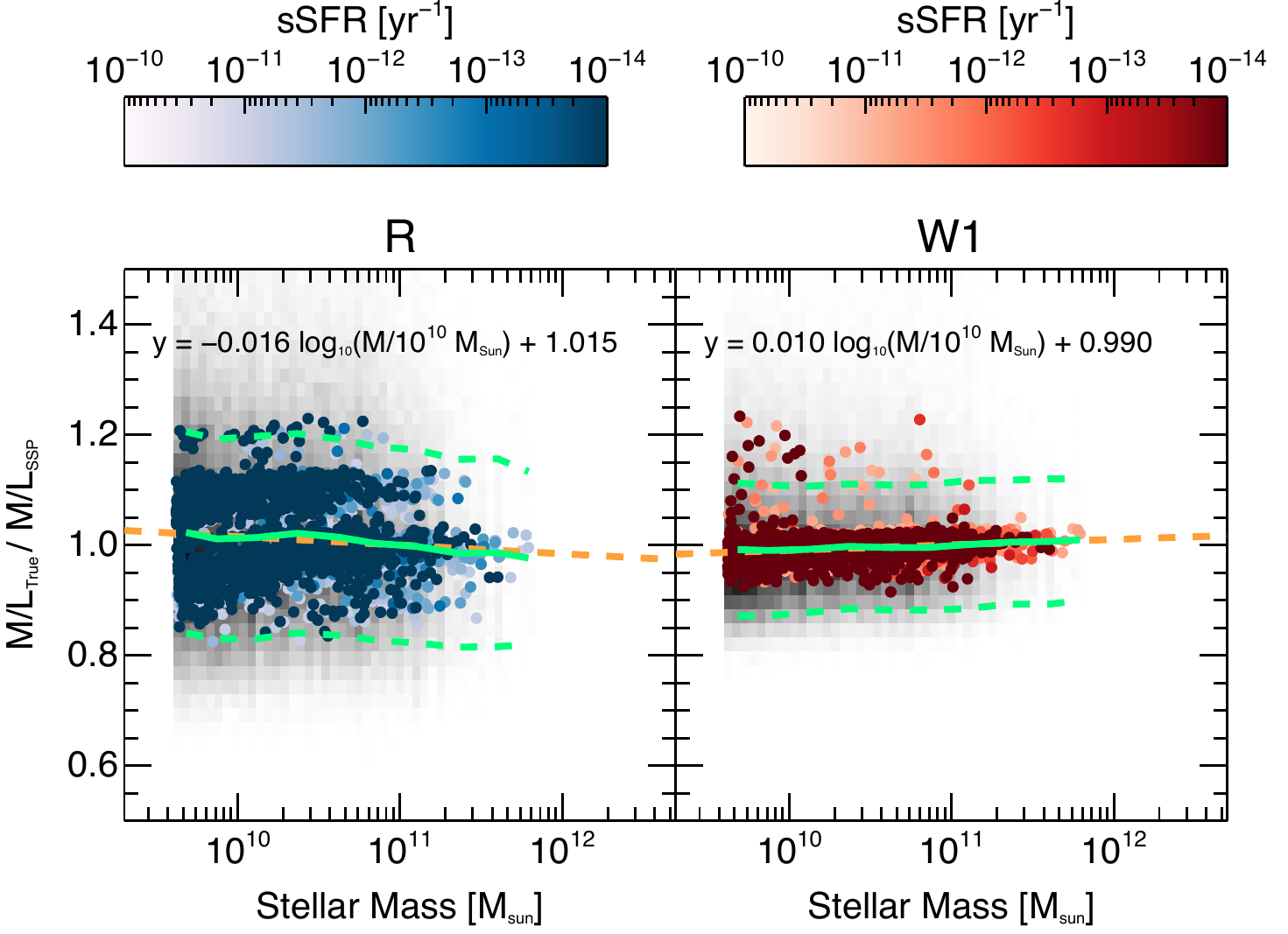}
   \end{turn} 
   \caption{The ratio of true M/L to that determined using the optically determined luminosity-weighted age and 
   metallicity to select the appropriate SSP M/L vs.\ total stellar mass (stars + remnants) for the full EAGLE 
   simulated galaxy sample. The left panel shows the result for the R band, the right panel for the W1 band. 
   The color scale in both cases indicates the sSFR for the simulated galaxies. In each panel the dashed orange 
   line is the best-fit linear trendline, this is the result of a robust linear fit with outlier rejection to the data 
   using the IDL code ROBUST$\_$LINEFIT.pro. The grey scale in the background indicates 
   the density of 50 Monte Carlo resimulations of each of the coloured data points with appropriate stellar population 
   errors (described in Section \ref{Sec:Effect_of_SFH}), the solid green lines are the median of the Monte Carlo 
   simulations and the dashed green lines are the $\pm$ 1-$\sigma$ uncertainties on the medians of the Monte Carlo
   simulations.}
   \label{fig:OffsetML}
\end{figure*}

Figure \ref{fig:OffsetML} shows the result of measuring the mismatch between the true M/L and the
luminosity-weighted SSP-equivalent M/L for each of the EAGLE galaxy SFHs. We plot the ratio of the 
true to SSP M/L vs.\ stellar mass to search for signs of a systematic trend in the mismatch parameter. 
Furthermore, the color scale indicates the sSFR of the simulated galaxies in order to search for signs 
of a correlation with this parameter. The orange dashed line is the best-fit linear relation to the colored dots 
(with the parameters of the fit indicated at the top of the panel). The grey scale shows the density of 50 
Monte Carlo simulations of the input data where to the luminosity-weighted age and metallicity 
noise was added following the prescription given in Section \ref{Sec:Effect_of_SFH}. The 
green solid line is the median of the grey points in bins of stellar mass, the green dashed lines are the 
$\pm$ 1$\sigma$ error bars on the median.

Several points are immediately apparent from an inspection of Figure \ref{fig:OffsetML}:
\begin{itemize}

\item The significant reduction in the scatter of the M/L$_{\rm True}$ to M/L$_{\rm SSP}$ ratio for the 
W1 relative to the R band is clear, especially before observational errors are applied (colored points). 
The scatter in the R band varies from 3$\%$ to 8$\%$, while in the W1 band it is only 1$\%$ to 3$\%$. 
This scatter represents the minimum possible uncertainty achievable when using this method to ascribe
a perfectly known luminosity-weighted average age and metallicity to a stellar population. 

\item In the W1 band there is a statistically significant systematic trend of increasing offset between true 
and SSP M/L with decreasing stellar mass. This trend appears to be due to the fact that the EAGLE 
simulations are able to reproduce the effect of cosmological downsizing \citep{Cowie96}. This
downsizing is the observation that, on average, lower-mass galaxies build up larger fractions of their stellar mass 
at lower redshift than high-mass galaxies. Again, where there is a larger fraction of younger stars, the 
SSP M/L has more difficulty accurately reproducing the true average M/L. The magnitude of this
offset is small, never more than 3-4$\%$ in the W1 and could be iteratively removed.

\item If such a mass dependent trend for the R band exists, it appears to be obscured by increased
scatter in that band due to its larger sensitivity to SFH. In contrast to naive expectations, the worst outliers in the R band
(those with M/L$_{\rm True}$ / M/L$_{\rm SSP}$ $>$ 1) are due to the \emph{oldest} galaxies. A quick glance at Figure 
\ref{fig:M_L_Grid} reveals the cause. The M/L ratio changes most dramatically at old ages; the outlying galaxies 
are those where most of their light comes from stars with a wide spread in metallicity at relatively old ages, typically
with little or no star formation later than 8 Gyr. For such galaxies the R band M/L of their stars can span
2 to 4.5 due to the relatively rapid period of metallicity increase common to most galaxies during the first 
$\sim$4 Gyr (see Figure \ref{fig:MassiveGalaxy}), no single SSP is able to adequately reproduce the M/L in this 
situation. Because these galaxies are then mostly truncated, they never get to build enough additional stars to
overpower this effect. In contrast for W1 the M/L change over the same metallicity range is 40-50$\%$ so the 
effect is much less pronounced.  

\item Finally, even when realistic observational errors are applied (see grey scale in the background), the 
1$\sigma$ scatter is a maximum of 13$\%$ in W1.
\end{itemize}

\begin{figure*} 
   \centering
   \begin{turn}{0}
   \includegraphics[scale=1.0]{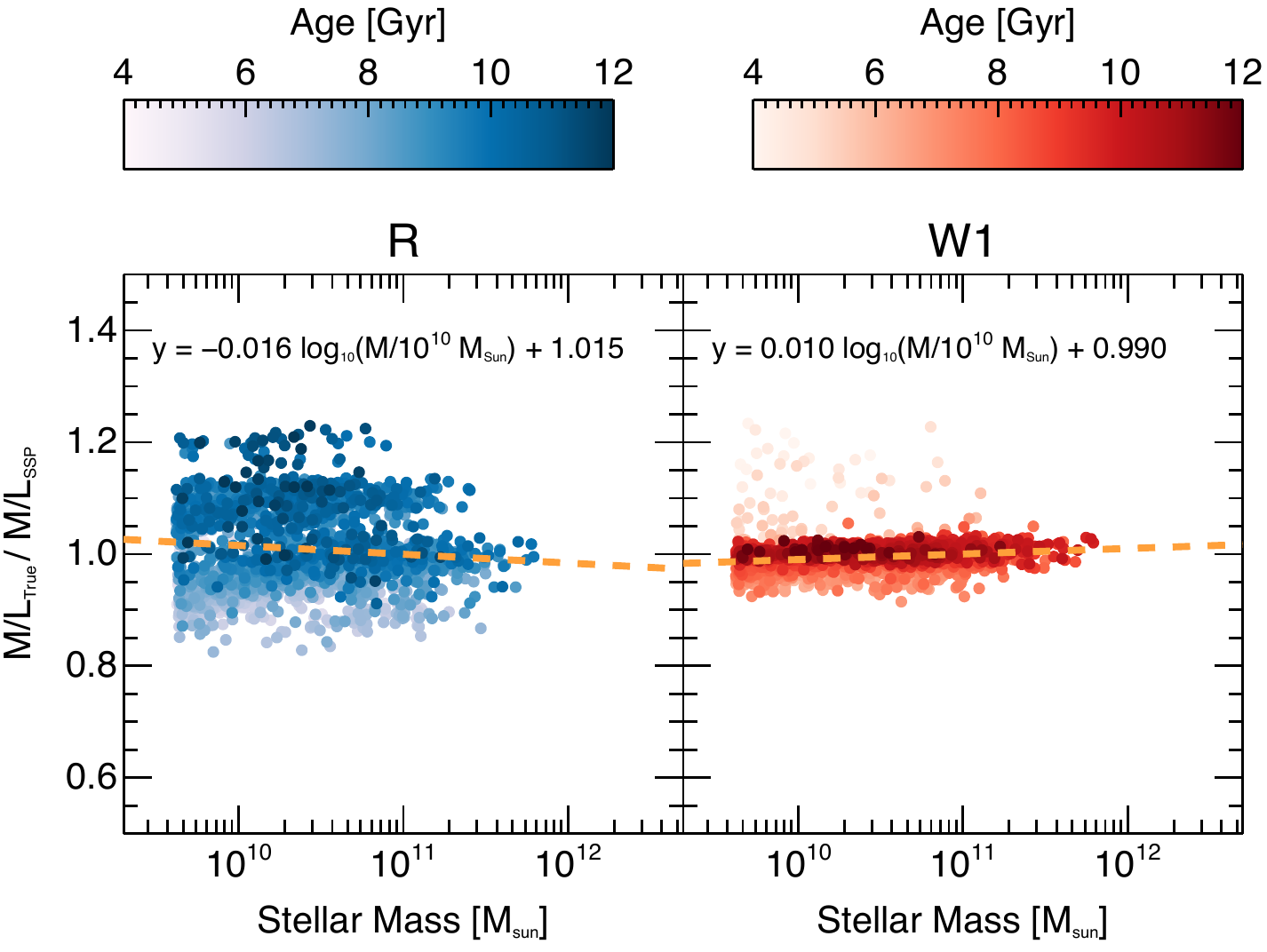}
   \end{turn} 
   \caption{This figure is the same as Figure \ref{fig:OffsetML} except that the color scale indicates the luminosity-weighted 
   age of the galaxy. It is clear that for galaxies with luminosity-weighted age $\ga$ 9 Gyr there is no systematic offset at all 
   in the W1 band i.e.\ the darker red points fall on true M/L to luminosity-weighted SSP-equivalent M/L = 1 and show no trend.}
   \label{fig:OffsetAge}
\end{figure*}

Figure \ref{fig:OffsetAge} again shows the mismatch between true and luminosity-weighted SSP-equivalent 
M/L against the stellar mass, only in this case the color scale indicates the luminosity-weighted age of each 
galaxy. We omit the additional Monte-Carlo simulations for clarity here. From this figure it is clear that offsets 
from the true to SSP M/L = 1 relation for W1 are driven by age differences. When the SSP equivalent age is $>$ 9 
Gyr, there is essentially no offset, no systematic trend with mass, and almost negligible scatter for the W1 
band (when neglecting observational errors).  We shall discuss this point further in Section \ref{Sec:discussion}.

\subsection{``Constant" Mass-to-Light Ratios}

In the previous sections we examined the use of the SSP-equivalent (i.e.\ luminosity-weighted) age and 
metallicity of a composite stellar population to derive the appropriate average M/L for that population. We 
found that given the current observational uncertainties this approach leads to a maximum systematic 
uncertainty (which can be removed) of only $\sim$4$\%$ and a random scatter (assuming the smallest
currently achievable observational uncertainties on SSP ages and metallicities)  of only 13$\%$. If the uncertainties 
on the observational measurement of SSP parameters could be entirely removed the theoretical 
minimum uncertainty introduced by differences in the SFH is a remarkably small $\sim$3$\%$ (for the most massive
galaxies).

We now examine the common situation whereby the spectroscopic data necessary for the inference of 
the SSP equivalent age and metallicity are unavailable. Previous studies have indicated that constant 
mass-to-light ratios in the NIR can be used to provide remarkably accurate stellar mass estimators for both 
early-type galaxies \cite[][]{Meidt14a} and even late type disks \cite[][]{Querejeta15,McGaugh15}. \cite{Meidt14a} 
used a range of exponentially declining SFHs combined with an empirically updated form of the \cite{BruzualCharlot} 
models to demonstrate that a single M/L ratio of 0.6 for the IRAC 1 3.6$\mu$m band can provide a good 
approximation to the true stellar mass, with only 0.1 dex scatter. Similarly, \cite{McGaugh15} used both a 
color-mass-to-light ratio relation from population synthesis models, and the Baryonic Tully-Fisher 
relation, to demonstrate that galaxy disks were well fit by a single M/L of 0.45 (at 3.6 $\mu$m) with a scatter of 
only 0.12 dex. Using the S$^{4}$G sample, \cite{Querejeta15} find essentially exactly the same M/L (M/Ls 
range from $\sim$0.4 to 0.55) for their sample of disk galaxies later than Sa. In fact, a M/L of 0.45 is consistent 
with the M/L predicted by our SSP models when the observed luminosity-weighted age and metallicity of spiral 
galaxy thin disks \citep[$<$ 1 Gyr in the outer regions of the disk;][]{Yoachim08} is used to select the appropriate 
M/L.

\begin{figure} 
   \centering
   \begin{turn}{0}
   \includegraphics[scale=1.0]{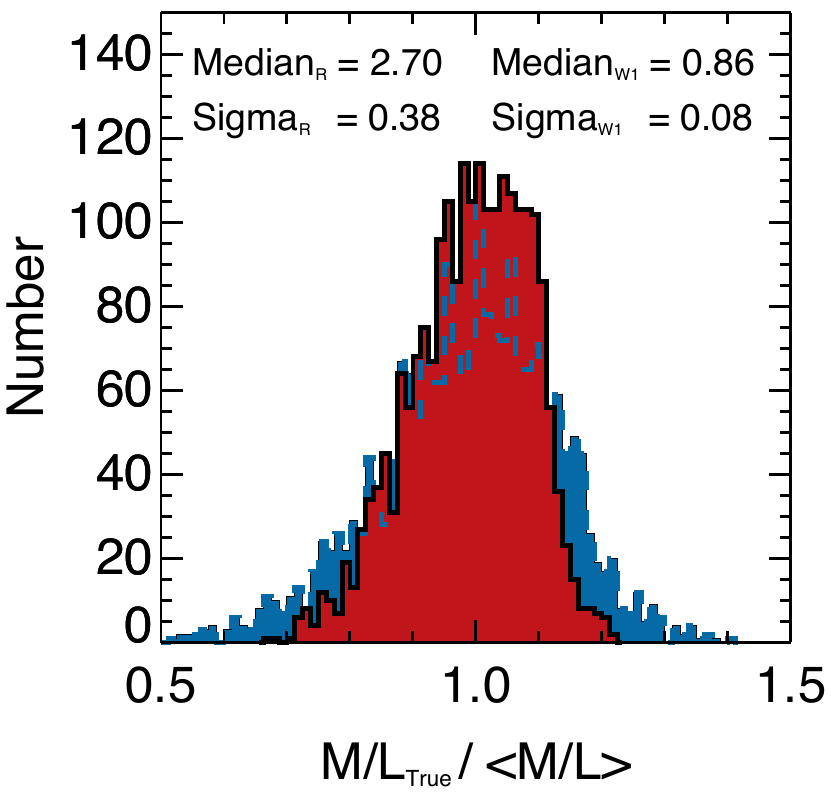}
   \end{turn} 
   \caption{Normalised histograms of the average mass-to-light ratios for the EAGLE sample 
   (defined in Section \ref{Sec:etg_selection}) in the R (blue histogram) and W1 (red histogram) 
   bands. The remarkably small scatter in the distributions is clear.  }
   \label{fig:EAGLE_ML}
\end{figure}

Our simulated EAGLE SFHs enable us to go a step further, and examine the M/L distribution of simulated 
quiescent/early-type galaxies. As Figure \ref{fig:EAGLE_ML} clearly demonstrates, the spread in M/L for the full 
sample is remarkably small, in the R band the median M/L is 2.70, with a 1-$\sigma$ scatter of only 0.38, 
for the W1 band the median M/L is 0.86 with a 1-$\sigma$ scatter of 0.08. The value of the median M/L 
is larger than the 0.6 found by \cite{Meidt14a}, which likely reflects the slightly different SFHs
assumed, slight bandpass and zeropoint differences between the IRAC 1 and W1 filters, and zeropoint 
differences between the two sets of SSP models used to determine the evolution of the M/L (a modified 
version of \citealt{BruzualCharlot} for \citealt{Meidt14a}, and \citealt{Bressan12,Bressan13} used here). 
Comparison with observed galaxies will be required before the exact zeropoint can be reliably determined
(i.e. by comparing dynamical and stellar masses derived with varying zeropoints). 
However, the general trends and behavior described here are independent of the zeropoint, and the results 
should therefore be robust, irrespective of the flavor of stellar population model used to construct the M/L grids.

The tight distribution of M/L displayed by the simulated quiescent/early-type galaxies indicates that using a single 
fixed M/L provides an accurate mass estimator, with scatter of only $\sim$9$\%$. This is only a factor of 
three larger than the theoretical minimum uncertainty of the SSP M/L method derived in Section \ref{RSFH}. 
It is also slightly better than the typical uncertainty ($\sim$13$\%$) of the SSP-based method, when the 
current typical observational uncertainties on the luminosity-weighted age and metallicity are taken into account. 

\begin{figure*} 
   \centering
   \begin{turn}{0}
   \includegraphics[scale=1.0]{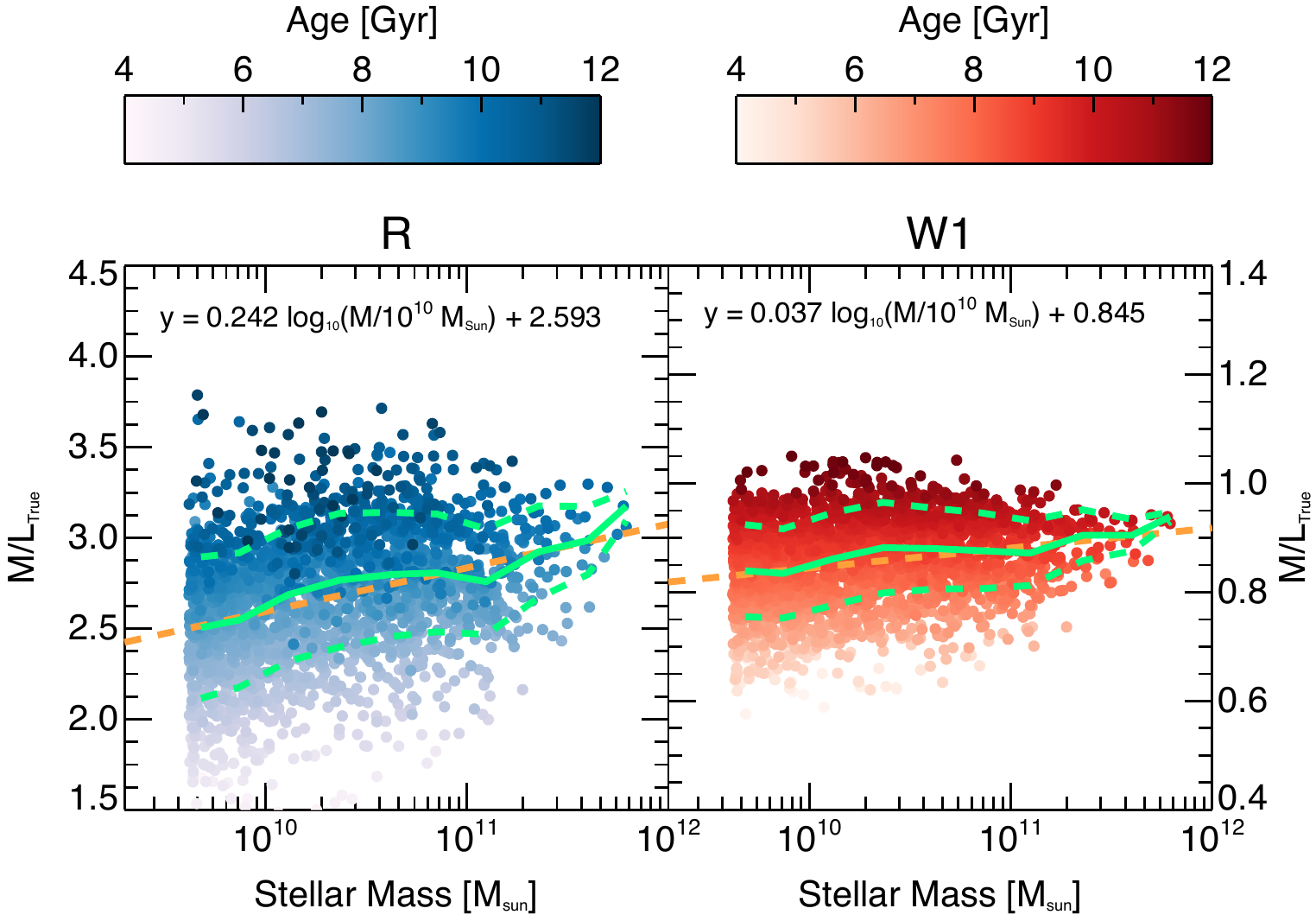}
   \end{turn} 
   \caption{The true M/L of each EAGLE galaxy plotted against stellar mass, symbols are the same as in the previous 
   figures. The mass dependence of the M/L is clear, as is the remarkably small scatter. The median M/L of the R and 
   W1 bands are 2.70 and 0.86 and can be used to iteratively determine an improved M/L (see text). We do not show 
   the SDSS matched sample as there is no significant difference. It is clear from this figure that the systematic change 
   in M/L with stellar mass is primarily driven by systematic changes in the average age of the galaxies with stellar mass.}
   \label{fig:EAGLE_ML2}
\end{figure*}

However, remarkably, it is possible to further improve the accuracy of this stellar mass indicator. As Figure 
\ref{fig:EAGLE_ML2} shows, the M/L correlates strongly with the stellar mass of the galaxy. This trend is 
driven primarily by the changing \emph{average} age of the galaxies with stellar mass, which itself is a result of the 
EAGLE simulations ability \citep{Furlong15} to reproduce a form of cosmological ``downsizing" \citep{Cowie96}. 
This is the observation that lower mass galaxies form more of their stars at later epochs
compared to higher mass galaxies, hence the average age of the low mass galaxies is lower than that of 
high mass galaxies.
The well-behaved nature of the trend, and the tightness of the relation (the scatter about the relation varies 
from a maximum of 9$\%$ to only 3$\%$) suggests that an iterative approach could be used to determine 
the appropriate M/L to use for any particular galaxy.

We have tested whether such an iterative approach will work and find that when starting with the median
M/L of the full population (2.70 in R and 0.86 in W1), and using the following fits:

\begin{eqnarray}
M/L_{\rm R}= 0.242 \times {\rm log_{10}}(M_\star/10^{10} M_\odot) + 2.593,
\end{eqnarray}

\begin{eqnarray}
M/L_{\rm W1}= 0.037 \times {\rm log_{10}}(M_\star/10^{10} M_\odot) + 0.845,
\end{eqnarray}

\noindent we find that convergence is obtained within 5 iterations for all of our EAGLE simulated galaxies. We discuss
in Section \ref{Sec:discussion} the 
implications and positive and negative aspects of both methods for determining the correct M/L in the 
discussion section.

\subsection{The Effect of Dust Extinction}

Despite the significant effect that relatively small amounts of dust extinction can have on derived stellar 
masses (especially in the optical), most studies of early-type galaxies do not attempt to measure or correct 
for extinction internal to the galaxy being studied (as opposed to Milky Way extinction). Instead, the majority 
of studies assume that early-types are dust free, despite many early-type galaxies being shown to have 
significant amounts of dust. For example, \cite{HCVSXVIII} find that around 17$\%$ of elliptical and 40$\%$ of lenticular galaxies 
in the Virgo cluster have measurable amounts of dust, a fraction that rises to 24$\%$ and 62$\%$ in a 
volume-limited sample of early-type galaxies \citep{HRS_ETGs}.

Here using the NIR bands to determine stellar mass again provides significant benefits. Based on 
the extinction corrections found in \citet{Schlafly11} and \citet{Yuan13}, the extinction in the W1 band 
is a factor of 12 less severe than in the R band. Although it must be noted that possible emission from 
the 3.3$\mu$m PAH feature can complicate exactly how much cleaner the W1 band is for a given dust 
mass. However, this effect is likely to be quite small, as \citet{Querejeta15} demonstrate that non-stellar 
emission always accounts for less than 15$\%$ of the IRAC1 emission for early-type galaxies from the S$^{4}$G 
sample. Given the R$_{\rm V}$ values from \citet{Schlafly11} and \citet{Yuan13}, an uncorrected uniform 
screen of dust providing 0.1 magnitudes of V band extinction would bias the M/L by 7$\%$ in 
the R band and by 0.6$\%$ in the W1 band. With 0.5 mag of V extinction the M/L would be incorrect by 
31$\%$ in the R and by 3$\%$ in the W1 bands respectively. Therefore for any reasonable amount of
dust extinction the effect would be negligible in the W1 band whilst being significant in the R band 
(and even worse in bluer bands).

\section{Discussion}
\label{Sec:discussion}

It is clear from our results that the NIR bands at 3.4 and 3.6 $\mu$m (and to a very similar extent 
the K band at 2.2 $\mu$m) provide an exceptional tracer of stellar mass. The previously noted reduced 
(relative to the optical) sensitivity of the M/L ratios of these bands to the age of a stellar population translates 
directly into a reduced sensitivity to the SFH as well.

Using the realistic SFHs from the EAGLE simulations we have demonstrated that when a fully resolved SFH 
is unavailable, the theoretical minimum uncertainty on a derived M/L is of the order of 3$\%$ 
when using the M/L of an SSP with the same luminosity-weighted age and metallicity as the stellar population. 
Though we caution that as discussed below the practically achieved minimum uncertainty is 
nearer 13$\%$ when typical errors on spectrally derived SSP age ($\sim$2 Gyr) are included, and the spectrally
derived SSP parameters are assumed to be unbiased compared to those of the true light-weighted parameters.
Furthermore, the theoretical minimum scatter does not account 
for the weak trend of a decreasing offset from unity of the ratio of the true M/L to SSP M/L with stellar mass 
observed in Figure \ref{fig:OffsetML}. This trend is driven by systematic 
changes in the population of galaxies with stellar mass, in the sense that at lower mass there are more young 
galaxies, and fewer uniformly old galaxies, while at high mass, there are essentially only old galaxies. This trend, 
which is a manifestation of cosmological downsizing, means that at all stellar masses the M/L of old galaxies is well 
reproduced by the SSP M/L, but galaxies which are younger (and found preferentially at lower mass) have more 
extended SFHs which are harder to reproduce with a single SSP equivalent M/L.

The minimum uncertainty is set by second order effects that are not fully captured by the use of a 
luminosity-weighted age and metallicity to ascribe an SSP M/L to a particular stellar population. In principle, using 
the observed residuals seen in Figure \ref{fig:OffsetAge} (e.g.\ the age dependence of the offset from the 1-to-1 
relation), it is possible to create an additional correction to further reduce this uncertainty. However, as Figure 
\ref{fig:OffsetML} amply demonstrates, this step would be largely irrelevant, as in practice the uncertainty on the 
SSP M/L in the NIR is dominated by the measurement uncertainties on the luminosity-weighted age and metallicity 
of the stellar population. For the parameterisation of the SSP age and metallicity uncertainties used here (chosen 
to match those found by the ATLAS$^{\rm 3D}$ survey), this uncertainty in the M/L is around 13$\%$. The 
typical uncertainties on the luminosity-weighted age and metallicity would have to be reduced by a factor of around 
3 before the magnitude of the two effects would become comparable, and the application of the additional age-dependent 
second-order correction would become advantageous.

Using the same simulated galaxy SFHs, we have also examined the efficacy of making use of ``constant" NIR 
M/L ratios to determine galaxy stellar masses. In common with other studies \citep{Meidt14a,McGaugh15}, we 
find that this approach provides a robust stellar mass. However, because of the extra information provided by 
our simulated SFHs, we are able to go a step further and employ an iterative approach to derive a mass-dependent 
M/L for a quiescent/early-type galaxy. This method accounts for the systematic variation of the population average M/L 
with galaxy stellar mass. This again is a manifestation of the cosmological downsizing, as Figure \ref{fig:EAGLE_ML2} 
demonstrates, the trend of decreasing M/L with decreasing stellar mass is driven by the decreasing average age of 
galaxies at lower stellar mass. We caution that the exact form of this correction depends on the ensemble 
properties of the simulated galaxies used to derive it, and the uncertainties on its use are likely to be significantly 
larger than the stated scatter, due in large part to a form of ``cosmic variance" and the relatively small number of 
simulated galaxies used to define the correction. For example, if we define the relation using quiescent/early-types selected 
exclusively in cluster environments, the average ages at all stellar masses would likely be higher and the slope of the 
relation would change somewhat. However, regardless of the exact form of the relation and hence 
on its applicability as a correction formula, we are confident that this 
behavior is an unavoidable consequence of the downsizing paradigm, and must be present in some form for real 
galaxy populations.

Despite the apparently improved accuracy provided by the mass-dependent M/L approach, we would still advise 
the use of the SSP-M/L approach in cases where the necessary SSP ages and metallicities are available. This is 
principally because the results of the SSP-M/L approach are more robustly model independent; the maximum 
systematic offset indicated by the models is only $\sim$3-4$\%$ over the mass range studied, and as long as the 
luminosity-weighted age of the galaxy is $>$ 4 Gyr, this uncertainty should not be exceeded. The accuracy of the 
mass-dependent M/L depends on how closely the sample of simulated galaxies used to derive it actually 
matches the population of galaxies it is being applied to. For example, if a large fraction of the observed galaxies 
are lower mass, but genuinely old (i.e.\ SSP age $>$ 9 Gyr), the mass-dependent M/L method would lead to a 
significantly larger error than expected.

In either case, the use of W1/IRAC1 (or K) photometry is strongly advised over the use of optical bands, simply 
because the contribution of dust extinction in quiescent/early-type galaxies is not zero and even 0.1 magnitudes of V extinction 
leads to a M/L bias of 7$\%$ in the R band, but less than 0.6$\%$ in W1.

\section{Conclusions}
\label{Sec:conclusions}

Using state-of-the-art simple stellar population models (SSP) and cosmological hydrodynamical simulations, 
we have examined the influence of the star formation history (SFH) and the metallicity distribution on our ability 
to recover accurate average stellar mass-to-light ratios in the optical and near infrared. We find that when 
additional information is available in the form of the luminosity-weighted (i.e.\ SSP equivalent) age and metallicity 
of the composite stellar population, it is possible to determine accurately the stellar mass of a quiescent/early-type galaxy 
using the NIR bands. We find that when using the W1 band as the stellar mass tracer for quiescent/early-type galaxies the 
maximum systematic uncertainty due to variations in the SFH is $<$4$\%$ for total stellar mass $>$ 10$^{9.5}$ 
M$_\odot$, with a SFH-dependent scatter at fixed stellar mass that is of a similar order. When assuming appropriate 
uncertainties on the luminosity-weighted age and metallicity ($\pm$ 2 Gyr when 12 Gyr old, $\pm$ 0.25 Gyr when 2 Gyr old, and 
uncertainty of 0.052 dex on metallicity at all ages), we estimate typical uncertainties on individual stellar
mass estimates to be $<$ 13$\%$ when using the W1 band, while the equivalent uncertainties for the R band
are 2-3 times larger even before internal dust extinction effects (which are negligible in the IR) are included. 
Where such high-quality, spectroscopically derived, SSP-equivalent ages and metallicities are 
available we suggest their use in conjunction with NIR W1/IRAC1 photometry as the most robust estimator of 
early-type stellar mass.

Furthermore, we present an iterative method for determining a mass-dependent M/L that yields formal
uncertainties $<$ 9$\%$. As this method requires no additional information, it is ideally suited for
large volume-limited surveys which match the environment distribution seen in the EAGLE simulations, but
which lack suitable spectroscopic observations to determine SSP-equivalent ages and metallicities. 

In cases where the particular galaxy survey is not likely to adequately sample the true volume-limited
galaxy population fairly (or more correctly the EAGLE representation of the population), in order to avoid introducing 
sample dependent correlations it would be preferable to assume a fixed NIR M/L.

In a forthcoming paper we will apply both these procedures to data from the ATLAS$^{\rm 3D}$ survey,
and use them to investigate the existence of a systematic variation in the IMF with stellar mass for 
quiescent/early-type galaxies.

\section{Acknowledgements}

The authors would like to thank the anonymous referee for their helpful comments
that greatly improved this paper.

RAC is supported by a Royal Society University Research Fellowship. We thank PRACE
for awarding us access to the Curie facility based in France at Tres`
Grand Centre de Calcul. This study used the DiRAC Data Centric
system at Durham University, operated by the Institute for Computational
Cosmology on behalf of the STFC DiRAC HPC Facility
(www.dirac.ac.uk); this equipment was funded by BIS National
E-infrastructure capital grant ST/K00042X/1, STFC capital grant
ST/H008519/1, STFC DiRAC Operations grant ST/K003267/1 and
Durham University. DiRAC is part of the National E-Infrastructure.
The study was sponsored by the Dutch National Computing Facilities
Foundation (NCF) for the use of supercomputer facilities,
with financial support from the Netherlands Organisation for Scientific
Research (NWO), and the European Research Council under
the European Unions Seventh Framework Programme (FP7/2007-
2013) / ERC Grant agreements 278594 GasAroundGalaxies, GA
267291 Cosmiway, and 321334 dustygal. Support was also received
via the Interuniversity Attraction Poles Programme initiated
by the Belgian Science Policy Office ([AP P7/08 CHARM]), the
National Science Foundation under Grant No. NSF PHY11-25915,
and the UK Science and Technology Facilities Council (grant numbers
ST/F001166/1 and ST/I000976/1) via rolling and consolidating
grants awarded to the ICC.

GvdV, and ES acknowledge support from the FP7 Marie Curie 
Actions of the European Commission, via the Initial Training Network DAGAL 
under REA grant agreement n$^\circ$ 289313.
SM acknowledges support from the German Science Foundation (DFG) via grant DFG SCHI 536/7-1.

\bibliographystyle{apj}
\bibliography{references}
\label{lastpage}

\end{document}